\documentclass[11pt, a4paper]{article}
\pdfoutput=1
\usepackage{graphicx}
\usepackage{jheppub}
\usepackage{lipsum}
\usepackage[all]{hypcap}
\usepackage{multirow}
\usepackage{titlesec}
\usepackage[T1]{fontenc}
\usepackage[utf8]{inputenc}  

\newcommand{\be}{\begin{equation}}
\newcommand{\ee}{\end{equation}}
\newcommand{\bear}{\begin{eqnarray}}
\newcommand{\eear}{\end{eqnarray}}
\newcommand{\ba}{\begin{array}}
\newcommand{\ea}{\end{array}}
 \usepackage{graphicx,tabularx}

\title{\large  The hunt for sub-GeV dark matter at neutrino facilities: a survey of past and present experiments}
\author[a,b]{Luca  Buonocore,}
\affiliation[a]{Dipartimento di Fisica, Universit\`a di Napoli Federico II and INFN, Sezione di Napoli, I-80126 Napoli, Italy}
\affiliation[b]{Physik Institut, Universit\"at Z\"urich, CH-8057 Z\"urich, Switzerland}
\author[c,d]{Claudia Frugiuele,}
\affiliation[c]{CERN, Theoretical Physics Departments, Geneva, Switzerland}
\affiliation[d]{INFN, Sezione di Milano, Via Celoria 16, I-20133
Milano, Italy.}

\author[e,f]{Patrick deNiverville}
\affiliation[e]{Center for Theoretical Physics of the Universe, IBS, Daejeon 34126, Korea}
\affiliation[f]{T2, Los Alamos National Laboratory (LANL), Los Alamos, NM, USA}

\emailAdd{luca.buonocore@na.infn.it}
\emailAdd{claudia.frugiuele@cern.ch}
\emailAdd{pgdeniverville@gmail.com}
\abstract{We survey the sensitivity of past and present neutrino experiments to MeV-GeV scale vector portal dark matter and find that these experiments possess novel sensitivity that has not yet fully explored. Taking $\alpha_D=0.1$ and a dark photon to dark matter mass ratio of three, the combined recast of previous analyses of BEBC and a projection of NO$\nu$A's sensitivity are found to rule out the scalar thermal target for dark matter masses between 10 MeV to 100 MeV with existing data, while CHARM-II and MINER$\nu$A place somewhat weaker limits. These limits can be dramatically improved by off-axis searches using the NuMI beamline and the MicroBooNE, MiniBooNE or ICARUS detectors, and can even begin to probe the Majorana thermal target. We conclude that past and present neutrino facilities can search for light dark matter concurrently with their neutrino program and reach a competitive sensitivity to proposed future experiments.}
\begin{document}

\maketitle
\section{Introduction}
\label{sec:intro}
A program for the direct detection of light dark matter (LDM) in the keV-GeV mass range has recently been advanced as many current dark matter searches are insensitive to DM below a few GeV in mass. This program has already borne fruit despite being only a few years old. It was shown that a new generation of DM direct detection experiments could be built with current or near-future technologies \cite{Essig:2012yx} and the first dedicated sub-GeV direct detection experiment (SENSEI) has already begun taking data \cite{Crisler:2018gci}.
It is timely to pose the question of how we can efficiently search for LDM in our laboratories. While high energy colliders have limited sensitivity to light, ultra-weakly coupled particles, accelerator experiments such as fixed-target experiments and low energy colliders (the so-called \textit{intensity frontier})  represent an ideal playground \cite{cosmicvision}, with the advantage that the DM is produced with relativistic energies \cite{Bjorken:2009mm,Batell:2009yf,Batell:2009di,Essig:2010gu}.
This has stimulated a wave of interest in accelerator-based LDM searches leading to the proposal of many new dedicated experiments  (e.g. SHiP \cite{Alekhin:2015byh}, LDMX \cite{ldmx1,ldmx2,ldmx3}, BDX \cite{bdx,bdx2}), which are under study by major laboratories.
The neutrino program is extensive, with many experiments currently running and even more in preparation, such as the Fermilab program at the Booster beamline with three liquid argon detectors: SBND, MicroBooNE, and ICARUS \cite{sbnd}.  
However, the attempt to make full use of existing neutrino fixed-target experiments for DM searches is limited to a few experiments,  analysis techniques  and DM signatures \cite{Batell:2009di, deNiverville:2011it, deNiverville:2012ij, Dharmapalan:2012xp, Batell:2014yra, Soper:2014ska, Aguilar-Arevalo:2017mqx, 
Aguilar-Arevalo:2018wea,beam1,beam2,beam3,nova,millicharged,trident,DeRomeri:2019kic}, with the strongest sensitivity coming from the NO$\nu$A \cite{nova} experiment at Fermilab \cite{nova}.  

In the present paper we will thoroughly investigate the potential of electron-DM scattering signatures at neutrino fixed-target experiments, considering for the first time the sensitivity of past and current experiments such as CHARM-II \cite{charm0}, BEBC  \cite{BEBC}, MINER$\nu$A \cite{Park:2015eqa} and MiniBooNE, MicroBooNE, and ICARUS as an off-axis detector for the NuMi beamline. 
The paper is organized as follows: in Sec. \ref{sec:Vec_Portal} we define a benchmark LDM model. Sec. \ref{sec:DM_prod} summarizes the main aspects of DM searches at neutrino facilities, and Sec. \ref{sec:Sensitivity} presents the results of the sensitivity studies.

\section{Vector Portal}
 \label{sec:Vec_Portal}
We consider as a benchmark model a dark sector coupled to the Standard Model through the vector portal. Specifically, we introduce a dark photon (DP) \cite{Holdom:1985ag}  $A'_{\mu}$ as the gauge boson of a new dark gauge group $U(1)_D$ kinetically mixed with the photon, 
and a scalar $\chi$ charged under $U(1)_D$ that serves as a DM candidate:
\be 
\mathcal{L}_{ \rm DM}=\mathcal{L}_{A^{\prime}}+\mathcal{L}_\chi 
\ee
where:
\be
\mathcal{L}_{A^{\prime}} =- \frac{1}{4} F'_{\mu \nu}F^{\prime \mu \nu} +\frac{m^2_{A'} }{2}A^{\prime \mu} A^{\prime}_{ \mu}-\frac{1}{2} \epsilon \;  F^{\prime}_{\mu \nu} F^{\mu \nu},
\ee
where $ \epsilon $ is the DP-photon kinetic mixing, while:
\be
\mathcal{L}_\chi =  i g_D  A^{\prime \mu} J_{\mu}^{\chi}+ \partial_{\mu} \chi^\dagger \partial^{\mu} \chi - m_{\chi}^2 \chi^\dagger \chi,
\ee
where $ J_{\mu}^{\chi}=  \left[ (\partial_\mu \chi^\dagger) \chi  -  \chi^\dagger  \partial_\mu \chi \right]$ and  $g_D$ is the $U(1)_D$ gauge coupling.
The region of the parameter space to which neutrino facilities are most sensitive is $ m_{A'} > 2 m_{\chi}$ and $ g_D \gg \epsilon e$, which implies that the DP  decays promptly into a $\chi\chi^\dagger$ pair. 
\par
We focus on the region where $\chi$ is a thermal relic compatible with the observed DM relic energy density. A complex scalar dark matter candidate $\chi$ is safe from constraints from precise measurements of the temperature anisotropies of the cosmic microwave background (CMB) radiation \cite{Lin:2011gj,Ade:2015xua}. Other compelling choices for DM  not in tension with the CMB includes a Majorana fermion or Pseudo-Dirac fermion with a mass splitting. In the following, we will also comment on these other candidates since the sensitivity of neutrino experiments to LDM does not significantly depend on its spin.
 
For $ m_{A'} > 2 m_{\chi}$, the annihilation cross section for a scalar dark matter particle can be written as \cite{Gordan}:
\be
\sigma (\chi \chi \to f \bar f) v  \sim \frac{8  \pi v^2 Y } { m_{\chi}^2}, 
\label{xthermal}
\ee
where $v$ is the relative DM velocity and $Y$ is defined as:
\be
Y \equiv \epsilon^2 \alpha_D \left (\frac{ m_{\chi}}{m_{A'}}\right)^4 ;
\ee
 we will present the sensitivity of neutrino facilities in the $ (Y,m_{\chi})$ plane since this allows us to identify the so called thermal targets, regions of the parameter space where, for a certain scenario, the correct DM thermal abundance is obtained \cite{Gordan,cosmicvision}. 
\par
 We consider as benchmark point $ \alpha_D=0.1 $ following \cite{PBC}, for which the most important existing constraints on the $ (Y,m_{\chi})$ are:
\begin{itemize}
\item \textbf{Laboratory bounds:}
the strongest laboratory constraints for $ m_{\chi} >60$\,MeV  come from a monophoton search performed by \textsc{BABAR} \cite{Lees:2017lec} that excludes the existence of a DP with $ \epsilon > 10^{-3}$ and $ m_{A'} < 8$ GeV decaying into $\chi\bar\chi$.
For a complex scalar with our benchmark parameters, \textsc{BABAR} bounds constrain thermal relics to be lighter than 100 MeV \cite{Lees:2017lec}.
The NA64 collaboration has recently published very strong limits for DP masses below 150\,MeV \cite{Banerjee:2017hhz} via a missing energy analysis. For large $ \alpha_D$, experiments looking at electron-DM scattering such as LSND \cite{lsnde,lsndpatrick}, MiniBooNE \cite{Aguilar-Arevalo:2017mqx,Aguilar-Arevalo:2018wea}, E137 \cite{Batell:2014mga,battaglieri1,battaglieri2} and NO$\nu$A \cite{nova} and capable of competing with NA64  for dark matter masses below few tens of MeV.

\item \textbf{Direct detection:} In the region where the $\chi$ relic abundance corresponds to the observed DM abundance and for large values of $\alpha_D$, CRESST-II and III place strong constraints on $m_\chi > 500$\,MeV \cite{Angloher:2014myn,Agnese:2013jaa,2012PhLB..711..264B,Petricca:2017zdp}.  However, as direct detection experiments lose sensitivity if DM is a Majorana or Pseudo-Dirac fermion, we will not present the constraints coming from direct detection in our sensitivity plots.
As was already mentioned in the introduction, many new ideas to probe the sub-GeV thermal DM parameter space via a direct detection experiment have been proposed \cite{cosmicvision}.
For example, SENSEI can discover or exclude the scalar thermal target for DM masses below 100 MeV \cite{Crisler:2018gci} in the near future, and Refs. \cite{Dror:2019onn,Dror:2019dib} detail a new fermionic dark matter signal that can potentially probe MeV-scale dark matter masses.

\item \textbf{Astrophysical and cosmological bounds:} The $U(1)$ gauge coupling $\alpha_D$ is bounded by the constraint on the DM self-scattering cross section coming from halo shape and bullet cluster observations, that is
\be
\frac{\sigma}{m_{\chi}} \lesssim \rm few \times cm^{2}/g.
\ee
In the whole MeV-GeV region values $\alpha_D \lesssim 0.1$ are allowed, while for $m_{\chi} > 10$ MeV even larger values of $\alpha_D $ up to
 $\alpha_D \lesssim 0.5 $ which is the upper bound suggested by the running of $\alpha_D$ \cite{Davoudiasl:2015hxa}. 
Furthermore, for the minimal DP model considered here a complex scalar lighter than 6.9 MeV is ruled out \cite{Boehm:2013jpa} by the Planck measurement of $N_\mathrm{eff}$ \cite{Ade:2015xua}.  
\end{itemize}

\section{DM production and detection at neutrino facilities}
\label{sec:DM_prod}

Fixed Target Neutrino facilities collide high-intensity proton beams with thick targets, producing large numbers of mesons whose leptonic decays generate a neutrino beam. The properties of the neutrino beam may be studied in both near and far detectors, located anywhere from tens of meters to hundreds of kilometers downstream of the target. Depending on the detector, both electron-neutrino and nucleon-neutrino interactions may be observed. Near detectors with relatively short baseline distances and large volumes can also serve as ideal LDM experiments \cite{Batell:2009di}. Rare meson decays (see Refs. \cite{Gninenko:2011uv,Gninenko:2012eq} for a previous approach to dark photon production through rare meson decays at the SPS for NOMAD, PS191 and CHARM-I) and bremsstrahlung can produce LDM alongside the neutrino beam mentioned above. These LDM particles can then be detected through their interactions with the nucleons and electrons of the neutrino detector, or if unstable and sufficiently long-lived, through their decays to visible particles. Electron scattering, in particular, provides one of the most promising signals for LDM particles with masses below 100 MeV \cite{lsndpatrick,nova,Aguilar-Arevalo:2018wea}.

The total number of DM particles produced through the decay of some pseudoscalar meson $\phi$ is given by:
\begin{equation}
 N_{\chi} = 2 N_\mathrm{POT} N_{\phi\mathrm{/POT}} \mathrm{Br}(\phi \to \chi \chi^\dagger)
\end{equation}
while the total number of DM particles produced in the target via bremsstrahlung is:
\be
N_{\chi }= \frac{ 2 N_{\rm POT}} {\sigma_{\rm T} (pp)} \sigma_{\rm T} ( p p \to A' X)
\label{DMtot}
\ee
where the factor of two takes into account the production of the  $\chi \bar \chi$ pair, $ N_{\rm POT}$ is the number of protons on target, and $ \sigma_{\rm T}(pp)$ is the total proton-proton cross section.
\subsection{ Electron-DM scattering inside the near detector}
DM-electron scattering is a very promising signature for new physics searches due the suppressed neutrino signal that can be further reduced with the appropriate cuts. 
We  can approximate  the inclusive electron-neutrino scattering cross section by \cite{Formaggio:2013kya}:
\be
\sigma(\nu_l e ) \sim 10^{-42} \left(\frac{E_{\nu}}{\rm GeV}\right) \rm cm^{-2}
\ee
while for $E_{\chi} \gg m_V$ the DM electron elastic cross section is:
\be
\sigma( \chi e) \sim \frac{4 \pi \alpha_D \alpha \epsilon^2 }{m_{A'}^2} \sim  10^{-27} \alpha_D \epsilon^2 \left (\frac{100 \; \rm MeV}{ m_{A'}} \right)^2  \rm cm^{-2}
 \ee 
 such that for $\epsilon \sim 10^{-4}-10^{-5}$ and a light DP the DM-electron scattering cross section is still orders of magnitude larger than the neutrino-electron cross section.
Hence we write the number of signal events $S_{\chi e \to \chi e } $  as:
\begin{equation}
S_{\chi e \to \chi e}  =  L_d n_{e} \; \int
dN_{\rm T}  (E_\chi)   \, \sigma( \chi e)  ~~.
\label{DIS1}
\end{equation}
where $n_{e}$ is the detector electron density , while
\begin{equation}
dN_{\rm T}  (E_\chi)  =  \, \epsilon_{\rm det}  \, N_{\chi } \,  \left(\frac{1}{\sigma} \frac{d\sigma}{ dE_\chi}\right)\!\! \left(pN\to \chi\bar{\chi}\right)_{\rm T} \; dE_\chi   ~~.
\label{fractionE}
\end{equation}
where $\epsilon_{\rm det}$ indicates the acceptance of the detector under investigation.
\par
It is challenging experimentally to distinguish an electron shower from the (large) neutral current events (NC) background.
However, elastic scattering events are characterized by no hadronic activity near the interaction vertex, and
\be
E \theta^2 < 2 m_e,
\ee
and indeed imposing this cut could reduce the NC background to a manageable level. This level of background reduction, however, requires detectors with good angular resolution.
A handful of experiments (LSND \cite{lsnde}, CHARM-II \cite{charm0,charm2}, MINER$\nu$A \cite{Park:2015eqa}, NO$\nu$A, MiniBooNE (MB) \cite{Aguilar-Arevalo:2018wea}  and MicroBooNE (MC) \cite{MC}) are equipped to distinguish such a signal. 
In the following section, we will study their sensitivity. Moreover, a new generation of liquid argon detectors will soon be running at Fermilab.
ICARUS is being installed and commissioned, and SBND is in the design and construction phase, and as such, we will also evaluate their future reach.
\begin{itemize}
\item \textbf{ CHARM-II  \cite{charm0,charm2}}  was a CERN based experiment which performed runs with proton energies of 400  GeV and 450 GeV. It took data from 1987 to 1991, collecting a total of $ 2.5 \times 10^{19}$ POT. 
The target calorimeter was 36 m long and consisted of 420 modules with cross sections of 3.70 $\times$ 3.70 m$^2$. The total detector mass was 692 tons with a fiducial mass of 450 tons (see Table \ref{tab:symbols} for important geometrical information). 
%Each module consists of a 4.8 cm (1/2X0) thick glass plate followed by a plane of 352 plastic streamer tubes with 1cm  spacing between the cathode wires.
CHARM-II performed a dedicated analysis of $\nu-e$ scattering \cite{charm2e}, which we can be recast to obtain its sensitivity to the sub-GeV DM parameter space. 

We took the number of $\pi^0$ ($\eta$) mesons to be 6.35$\times$POT (0.726$\times$POT), with their momenta and angular distribution determined by a PYTHIA 8 simulation (see sec. \ref{ssec:sim} for further details). We selected dark matter-electron scattering events with electron recoil energies between 3 and 24 GeV and assumed a reconstruction efficiency of 0.73. We placed a 90\% limit on 340 dark matter induced electron recoil events.

\item \textbf{ BEBC/WA66  \cite{BEBC}}  The WA66 experiment used the Big European Bubble Chamber (BEBC), a large detector located at CERN and installed in the early 70's, to detect neutrinos produced by dumping 400 GeV protons from the CERN SPS into a copper block large enough to contain almost the entire hadronic cascade. This long target suppresses the standard neutrino flux by almost three orders of magnitude (i.e., emitted by pions or kaons decay), while prompt neutrinos (for instance those created by $D-$meson decays) were still copiously produced and reached the detector.
This specific feature makes this experiment suitable for new physics searches, and hence a new physics analysis is available to be recast \cite{bebcSUBIR}. 

BEBC used a lower energy beam than CHARM and produced slightly fewer mesons as a result, with $N_{\pi^0}=6.15\times$POT and $N_\eta=0.703\times$POT. The analysis cuts used were: $E \theta^{2}< 2 m_e$,   $E_\mathrm{min }^\mathrm{reco} > 8 $ GeV  with a reconstruction efficiency of 0.8. The $90\%$ confidence limit corresponds to 3.5 new physics events.
\item  \textbf{ NO$\nu$A  \cite{Park:2015eqa}} is a Fermilab-based long-baseline neutrino experiment located slightly off-axis from the NuMI beam. Its near detector is located 990 meters downstream of the NuMI target with 125 tons of active mass. The reach of the existing  neutrino-electron analysis \cite{biao} was previously studied in Ref. \cite{nova}.
The  following cuts are applied:  $E \theta^{2}< 5  \; \rm{MeV}~\rm{rad}^{2}$  and the recoil energy is considered in the range 0.5 GeV-5 GeV. The reconstruction efficiency was taken to be 50\% with a total background of $ \sim 580$ events for $ 2.97 \times 10^{20}$ POT \cite{nova,biao}.
\item \textbf{ MINER$\nu$A  \cite{Park:2015eqa}}  is a neutrino scattering experiment currently running that uses the NuMI beam-line at Fermilab. 
It performed a neutrino electron scattering analysis \cite{Park:2015eqa} intending to improve the precision in measuring the neutrino flux. However, possible new physics contamination from DM electron scattering was not taken into account. We will study here for the first time whether this contamination might be significant.
The number of mesons produced by the NuMI beamline was estimated by PYTHIA to be $N_{\pi^0}=4.176\times$POT and $N_\eta=0.474\times$POT. We applied the following cuts $E \theta^{2}< 3.2 \; \rm{MeV}~\rm{rad}^{2}$ in our analysis, and placed a 90\% exclusion on 41 dark matter induced recoil events.
\item \textbf{ MiniBooNE off-axis (MBOA)} is a Fermilab-based 800 ton detector.  It collected data both as an on-axis detector from the Booster Beamline (8.9 GeV) and as a detector located $6.5^\circ $ off-axis from the NuMI beamline (120 GeV) \cite{MBoff}. An analysis considering DM-electron scattering was recently published by the MiniBooNE collaboration \cite{Aguilar-Arevalo:2018wea} considering an 8.9 GeV run in beam dump mode.
Here we consider instead the possible sensitivity of the off-axis NuMI data with the same meson production estimates as those quoted for MINVER$\nu$A above.  We applied the same cuts as \cite{Aguilar-Arevalo:2018wea}  ($\cos{\theta} >0.99,$ $E_\mathrm{min }^\mathrm{reco} > 75 $ MeV, assumed a reconstruction efficiency of 0.35 and consider a background free analysis, as an off-axis signal should have greatly reduced beam related backgrounds. 
\item \textbf{MicroBooNE off-axis} MicroBooNE is the first large liquid-argon time projection chamber (LArTPC) to acquire a high statistics sample of neutrino interactions. It is located at a $7.5^\circ$ angle relative to the NuMI beamline. We consider the same cuts and production rates as MiniBooNE off-axis.
\item \textbf{ICARUS off-axis} ICARUS is a 600 ton (500 ton fiducial) LArTPC that serves as the far detector of the SBND program. It is located at a $5.7^\circ$ angle relative to the NuMI beamline. We consider the same cuts and production rates as MiniBooNE off-axis.
\end{itemize}

\begin{table}[h]
\begin{tabularx}{\textwidth}{ |X|X|X|X|X| }
  \hline
    {\bf Experiment }& {\bf $d$ (m)} &{\bf  $n_{\rm det}$} (g/cm$^3$)& {\bf  Mass (tons)} &{\bf POT} \\
  \hline
 
 \hline
  MB NuMI \cite{MBoff}   120 GeV&  745 &  0.69 & 800& 6 $ \times 10^{20} $\\
\hline
  MC NuMI \cite{MC}   120 GeV&  684 &  1.4 & 89 & $  10^{21} $\\

\hline

MINER$\nu$A  ~\cite{Park:2015eqa} 120 GeV  & 980   & 0.9  &6.1  &$3.43\times10^{20}$ \\
 
 \hline
 
 CHARM-II ~\cite{charm0} 450 GeV  & 480  & 1.3 & 692 &$2.5 \times10^{19}$ \\
\hline
 BEBC/WA66 ~\cite{BEBC} 400 GeV  & 480  & 0.69  & 11.5  &$2.72 \times10^{18}$ \\
 \hline
ICARUS NuMI ~\cite{Amerio:2004ze} 120 GeV & 789 & 1.4 & 500 & $10^{21}$ \\
  \hline
\end{tabularx}
%}
\\
\caption{Summary of experiments and their geometry. POT stands for the total number of protons on target, BE the energy of the proton beam hitting the target, and  $d$ indicates the distance of the detector from the target.  The BEBC, CHARM-II, and MINER$\nu$A detectors are located on-axis with respect to the beamline, while MB, MC, and ICARUS are located off-axis by an angle of $\theta = 6.5^ \circ, 5.7^\circ$, and $7.5^\circ$ from the NuMi beamline, respectively.}
\label{tab:symbols}
\end{table}

\subsection{Simulation of the signal}
\label{ssec:sim}
In the parameter space relevant for fixed target neutrino experiments, the generation of signal events can be modeled as a three-step process: 
\begin{enumerate}
\item (prompt) production of dark matter particles in the target or proton beam dump;
\item propagation (as free particles) from the production point to the detector; 
\item interaction within the active volume of the detector.
\end{enumerate}
The production rate of DM particles is dominated by the interaction of the incoming protons within the first few interaction lengths in the dump,
with the most relevant mechanisms given, as mentioned above, by prompt radiative meson decays and proton bremsstrahlung.
We neglect effects related to the geometry of the production target (and secondary particle interactions),
as its characteristic length is far smaller than the distance between the beam dump and the detector, and we assume that the production is localized to a point at the center of the target.
The simulation of the full production and propagation process was performed using two different available tools, BdNMC~\cite{deNiverville:2016rqh} and MadDump~\cite{maddump}.
They both provide a complete framework to handle all the three-particle generation steps in a transparent and mostly-automatic fashion.

Nonetheless, the two tools differ in many aspects regarding their actual implementation, providing a powerful test of the robustness of our prediction. In particular, they handle the DM scattering process inside the detector (step 3) following two different strategies. BdNMC works event-by-event and decides if each DM particle reaching the detector will interact according to an acceptance-rejection criterion. If an event is rejected, a new one is generated, and the
procedure is iterated until the requested number of sample events is reached.  In MadDump, the intermediate results of step 1 and step 2 are used to build a fake DM beam, characterized by its bidimensional flux distribution in energy and angle, which interacts within the detector acceptance. In this way,
the interaction probability (cross section) can be computed by exploiting standard Monte Carlo methods, and the final signal events can be generated through an efficient unweighting procedure (as provided by the MadGraph framework~\cite{Alwall:2011uj}).

Limiting our focus to the cases relevant to this work, we have found a reasonably good agreement, within a few percent, between the predictions of
BdNMC and MadDump on the total signal rates with and without applying the selection cuts on the electron recoil.
The level of agreement is below the main experimental and theoretical systematics. One of the primary sources of uncertainty
is given by the modeling of the meson spectra produced in the proton dump, which represents an input for our tools.
Indeed, BdNMC and MadDump handle only the decay of the mesons into DM particles within an effective field theory approach.
External data must be supplied, and one can either rely on full event-generator such as PYTHIA~\cite{pythia8} or adopt a phenomenological parametrization such as those provided in Ref.~\cite{Bonesini:2001iz},
which represent the default choice in BdNMC. We have found that for the relatively high energy beams of the neutrino experiments
investigated in this work, the difference in the final rates can be as large as a factor of two, with the distribution given by PYTHIA
being softer and with a larger angular spread. We assume a pragmatic approach adopting the more conservative result given by PYTHIA,
which has been investigated in Ref.~\cite{Dobrich:2019dxc}.

\begin{figure}
 \centerline{\includegraphics[width=0.8\textwidth]{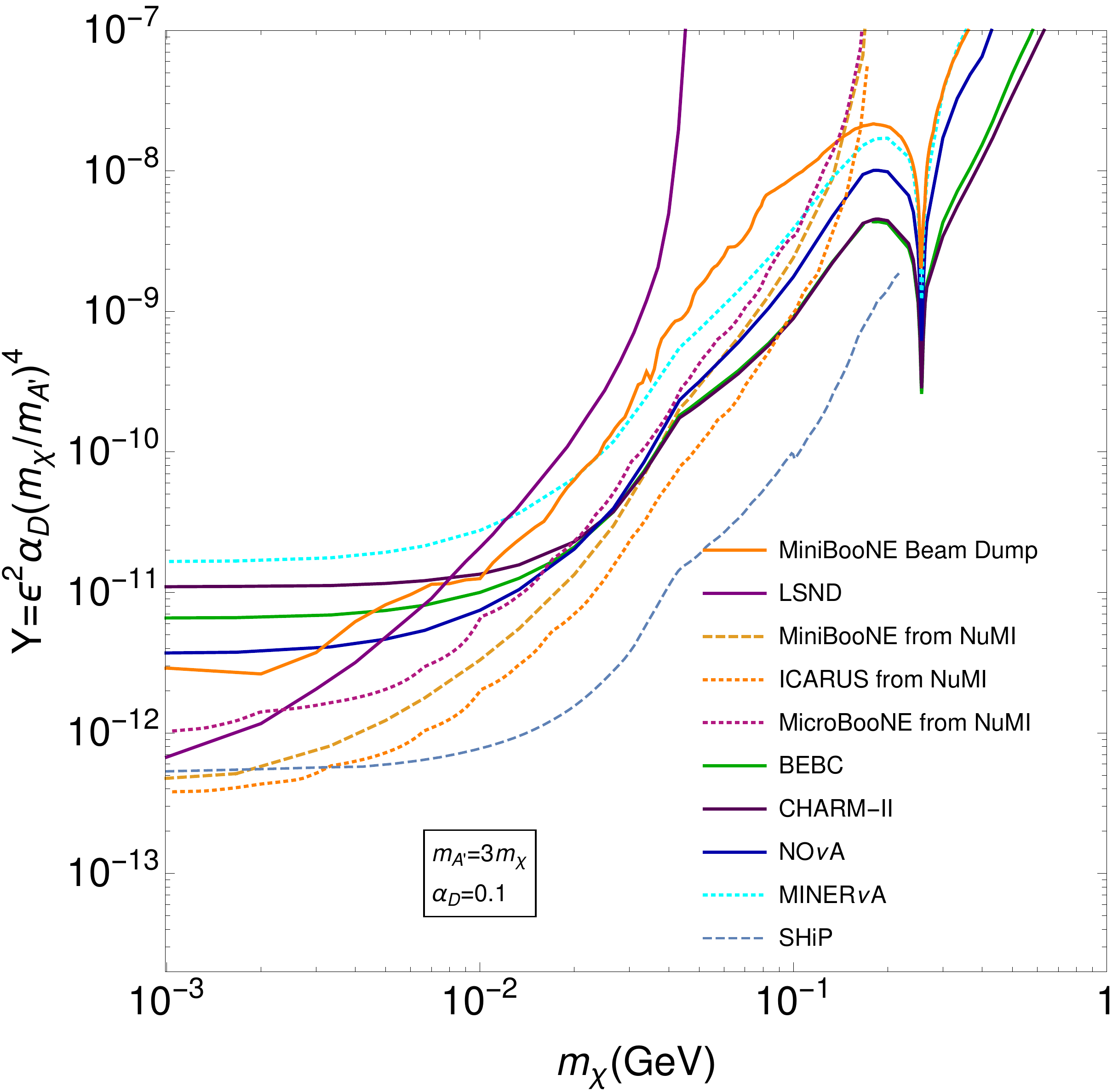}}
 \caption{All limits and projections from existing fixed target neutrino experiments. Limits based on existing data and analyses are given by solid lines, while projections are dotted.}
 \label{fig1}
\end{figure}

\section{Sensitivity to sub-GeV DM of past and current experiments }
\label{sec:Sensitivity}
\begin{figure}
\centerline{\includegraphics[width=0.8\textwidth]{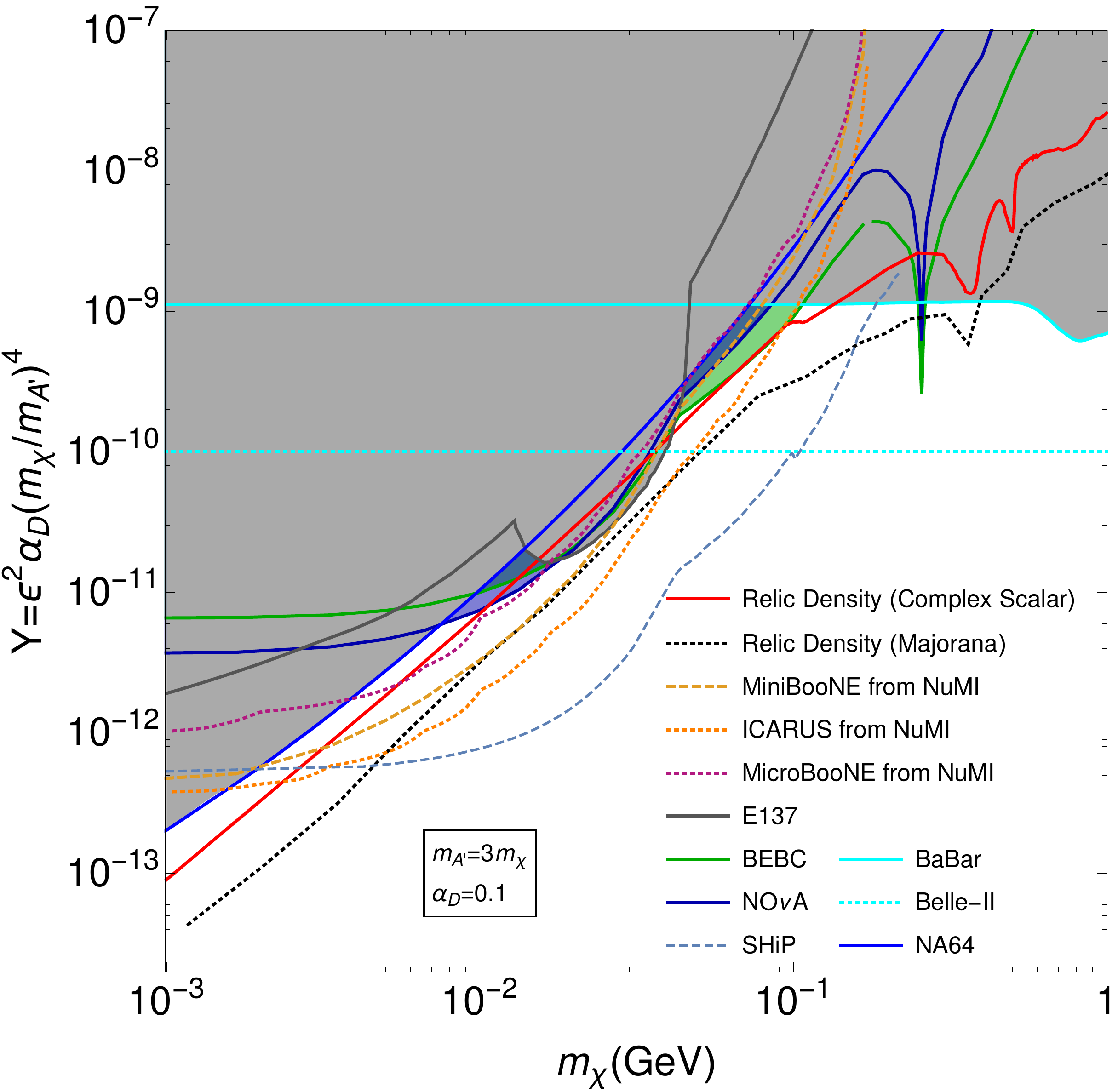}}
\caption{We show a slice of the vector portal dark matter parameter space with $\alpha_D=0.1$ and $m_V = 3 m_\chi$. The solid (dotted) black lines show the parameter space for which a complex scalar (Majorana) dark matter candidate coupled to a DP reproduces the observed dark matter relic density. The blue shaded region is excluded by the NO$\nu$A experiment, while the gray shaded region is excluded by a recast of a physics analysis of BEBC. The other dotted lines show the projected sensitivity of a new physics analysis of $10^{21}$ POT of data for MiniBooNE, ICARUS, and MicroBooNE taking data from the NuMI beamline. SBN is too far off-axis to provide much sensitivity to vector portal dark matter produced by the NuMI beamline, and is not shown.}
\label{fig2}
\end{figure}

In Fig. \ref{fig1} we present the comparison of the sensitivity of all different neutrino experiments described above, including also previous results such as  NO$\nu$A \cite{nova} and MB on-axis \cite{Aguilar-Arevalo:2018wea}, while in Fig. \ref{fig2}  we compare the strongest ones to existing constraints described in Section 2.
\par
We find that:
\begin{itemize}
\item In the small mass region $m_{\chi} \lesssim 50 $ MeV, the best sensitivity is reached by MB off-axis, which can rule out part of the thermal targets both for scalar and Majorana DM. NO$\nu$A is capable of excluding some parameter space for $m_\chi \approx 10\,\mathrm{MeV}$.  ICARUS will further improve on this result reaching a sensitivity to the Majorana target even better, while
MicroBooNE has more limited reach, although the off-axis run still could improve over the beam dump dark matter run using the 8.9 GeV Booster Beamline. Both the MicroBooNE and ICARUS analyses assume zero background based upon the results of the MiniBooNE electron scattering analysis, but this may be too optimistic. MicroBooNE and ICARUS use different detector medium and technology than MiniBooNE, and may not be able to attain the same level of background rejection as MiniBooNE was capable of during its beam dump dark matter search. Conversely, the off-axis position considered should greatly reduce the number of neutrinos reaching the detector, which may improve the potential sensitivity of all three experiments. Similar sensitivity could potentially be achieved by repeating the MiniBooNE-DM beam dump run \cite{Aguilar-Arevalo:2018wea} with the ICARUS and MicroBooNE detectors but we have not performed a full analysis for this work.
\item For higher masses, the best reach amongst fixed target experiments instead comes from old SPS experiments like BEBC and CHARM-II. In particular, the recast of the previous new physics search using BEBC \cite{bebcSUBIR} eliminates some existing parameter space not covered by E137, NA64, and \textsc{BABAR}. 
\item  MINER$\nu$A is less sensitive to new physics than other existing experiments, but for sufficiently high mass, it can surpass the sensitivity of the MiniBooNE beam dump search \cite{Aguilar-Arevalo:2018wea}. However, both NO$\nu$A  and the old SPS experiments (CHARM-II and BEBC) have a significantly better reach.
\end{itemize}

\section{Conclusions}
In this paper, we surveyed the reach of past and present neutrino facilities. We found that:
\begin{itemize}
\item NO$\nu$A and BEBC exclude a significant range of masses for the scalar thermal target. A dedicated DM analysis by NO$\nu$A is important, as it could further improve on this result.
\item An analysis performed on the existing data of MB from the NuMi beam could rule out most of the remaining parameter space and even reach the Majorana thermal target, substantially improving on the reach of the MB beam dump dedicated run. However, as such an analysis may not occur, it is critical that the potential of existing and future experiments such as MicroBooNE and ICARUS be exploited. We also find that the signal improves as the threshold for the electron recoil energy is decreased, a trait that could be targeted by future analyses.
\item ICARUS rules out the Majorana thermal target for masses between 6 and 50 MeV. This result is highly complementary to Belle II and not far from the reach of SHiP \cite{PBC}, as shown in Fig. 2. We limit our off-axis analyses to MicroBooNE and ICARUS, as SBND was found to be too far off-axis to achieve good acceptance.
\item Our final conclusion is that existing and past facilities can compete with future and proposed experiments sensitivity \cite{PBC} in a completely parasitic way to their neutrino program. 
\end{itemize}

\begin{acknowledgments}
We are indebted to Subir Sarkar for informing us of the BEBC analysis and helpful discussions regarding its details, and to Brian Batell for valuable feedback. The 
work of PdN was supported in part by IBS (Project Code IBS-R018-D1) and 
Los Alamos National Laboratory under the LDRD program.
\end{acknowledgments}

\newpage

\bibliographystyle{JHEP}
\bibliography{scatteringTOT}

\end{document}